\documentclass{phbauth}        
\usepackage{graphicx}

\begin{document}

\begin{frontmatter}

\title{Deformation of Fermi Surface Due to 
Antiferromagnetic Correlation}

\author[address1]{K. Morita\thanksref{thank1}},
\author[address1]{K. Miyake},

\address[address1]{Department of Physical Science, \\
Graduate School of Engineering Science, Osaka University, \\
Toyonaka, Osaka 560-8531, Japan }

\thanks[thank1]{Corresponding author. E-mail: morita@eagle.mp.es.osaka-u.ac.jp} 

\begin{abstract}

It is investigated, on the basis of the fluctuation exchange approximation,
how the shape of the Fermi surface is modified 
in two-dimensional $t-t'-U$ Hubbard model 
at half filling as strength of the onsite Coulomb interaction $U$ is 
increased. The antiferromagnetic (AF) correlation length is shown to be 
enhanced as the Coulomb interaction get closer to the critical value 
$U_{\rm c}$ for the onset AF order.  At the same time, the 
shape of the renormalized Fermi surface is deformed showing the tendency of 
nesting near $U_{\rm c}$. 

\end{abstract}


\end{frontmatter}




Layered cuprates have attracted much attention over the past decade
because they exhibit anomalous behaviors
in the normal state 
except for the fact that they have
record high transition temperature ($T_{\rm c}$). 
Some of the anomalous properties
have features expressed 
by keywords, spin-charge separation aspect, spin-gap behavior,
and so on
\cite{ex2,ex5,scsepa1}.
They cannot be explained by the conventional Fermi liquid theory
in its simple form.   
It is, however, expected that the Fermi surface of quasi particle
becomes nearly nested
near the antiferromagnetic (AF) insulator phase
\cite{miyake3}.
Then it has been shown
that such anomalous properties can be understood 
as effects of AF spin fluctuations
with technically nested Femi surface
\cite{miyake1,miyake2}.
Yanase-Yamada calculated 
the one-particle self-energy on the one loop approximation
using phenomenological form of
the spectrum of magnetic excitations,
and pointed out that the strong AF spin fluctuations
works to deform the Fermi surface as to be nested
\cite{yanase}.
This problem has also been studied
in 2D Hubbard model on the square lattice
by the self-consistent second order perturbation theory (SCSOPT) 
\cite{nojiri},
and in 2D Hubbard model on the triangular lattice 
by a simple second order perturbation
\cite{yoda}.
These theories, however, cannot treat of spin fluctuations properly.
Because of that, only a little tendency of change could be observed.
The purpose of this paper is to study,
on the basis of the fluctuation exchange (FLEX) approximation,
how the shape of the Fermi surface is modified 
in  two-dimensional $t-t'-U$ Hubbard model at half filling,
which is one of the simplest model having features of 
layered high-$T_{\rm c}$ cuprates, 
as strength of $U$ is increased.
By adopting FLEX
\cite{FLEX}, 
we can take into account an effect of strong AF spin fluctuations.

The Green function and self-energy on the real axis 
are calculated by the standard method
\cite{calc1,calc2,calc3},
and the Fermi surface and chemical potential are determined
so as to satisfy the Luttinger sum rule
\cite{agd}.
We have retained $128\times128$ lattice points
and $512 (\equiv N)$ discrete points of energy.
Both of the cut-off value of energy $\varepsilon_{\rm c}$
and Matsubara frequency $\varepsilon_{\rm nc}=(2N-1)\pi T$
are $30t$ corresponding to the temperature $T/t=0.0093$.
A result of deformed Fermi surface is shown in Fig.~1
for the system with $U/t=3.12$, $t'/t=0.2$, 
$t$ and $t$ being the transfer between the nearest neighbors 
and the second nearest neighbors.
The AF correlation length $\xi/a$,
$a$ being the lattice constant,
is estimated as $\xi/a=14.8$
on FLEX approximation.
The degree of deformation of the Fermi surface
at the zone boundary $k_{y}=\pi$ is
$\Delta k_{x}=-0.067 \pm 0.025$
and along the line $k_{x}=k_{y}$ is
$\Delta k_{\perp}=0.017 \pm 0.017$
for $U/t=3.12$,
where $\Delta k$ is the difference between
the Fermi wave number without interaction
and the renormalized one. 
As we can see in Fig.~1, the Fermi surface is
deformed in the direction to the nesting.
The corresponding values of $\Delta k$ for $U/t=2$ is
$\Delta k_{x}=-0.018 \pm 0.025$ and 
$\Delta k_{\perp}=0.017 \pm 0.017$,
which can hardly be seen if they are drawn as Fig.~1.

Deformation of the Fermi surface, calculated
by SCSOPT in 2D Hubbard model near the half-filling,
cannot be seen clearly even for rather large value of interaction $U/t=4$
\cite{nojiri}.
This is also the case in Ref
\cite{yoda}.
On the other hand,
change of the Fermi surface we calculated is not so large as 
that of Yanase-Yamada who claimed that 
much larger deformation is obtained even smaller value of $\xi/a=6$
\cite{yanase}.

We acknowlege O. Narikiyo for fruitful discussions.
One of us (K. Morita) is grateful to
Y. Okuno and T. Takimoto 
for their useful comments and advice about 
a technique of numerical calculation.
This work is supported by a Grant-in-Aid for COE Research (10CE2004)
of Ministry of Education, Science, Sports and Culture.

\begin{figure}[t]
\begin{center}\leavevmode
\rotatebox{0}{\includegraphics[width=0.9\linewidth]{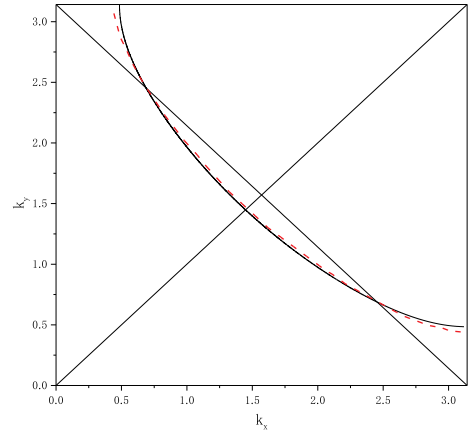}}
\caption{Deformation of Fermi surface.
Solid line is for the Fermi surface without interaction,
and dashed line for the renormalized Fermi surface.}

\end{center}\end{figure}


\end{document}